\documentclass[sigconf,nonacm]{acmart}
\AtBeginDocument{%
  }

\setcopyright{acmlicensed}
\copyrightyear{2018}
\acmYear{2018}
\acmDOI{XXXXXXX.XXXXXXX}
\acmISBN{978-1-4503-XXXX-X/2018/06}

\usepackage{enumitem}
\usepackage{graphicx}
\usepackage{subfig}
\usepackage{multirow}

\begin{document}

\title[Social-Physical Interaction with Virtual Characters]{Exploring Immersive Social–Physical Interaction with Virtual Characters through Coordinated Robotic Encountered-Type Contact}

\author{Eric Godden}
\orcid{0009-0002-2345-8770}
\affiliation{%
  \institution{Queen's University}
  \city{Kingston}
  \country{Canada}
}
\email{18eg16@queensu.ca}

\author{Jacquie Groenewegen}
\affiliation{%
  \institution{Queen's University}
  \city{Kingston}
  \country{Canada}}
\email{21jng3@queensu.ca}

\author{Michael Wheeler}
\affiliation{%
  \institution{York University}
  \city{Toronto}
  \country{Canada}
}
\email{mwheeler@yorku.ca}

\author{Matthew K.X.J. Pan}
\affiliation{%
 \institution{Queen's University}
 \city{Kingston}
 \country{Canada}}
\email{matthew.pan@queensu.ca}


\begin{teaserfigure}
  \includegraphics[width=\textwidth, trim=0cm 9.5cm 0cm 9.5cm, clip]{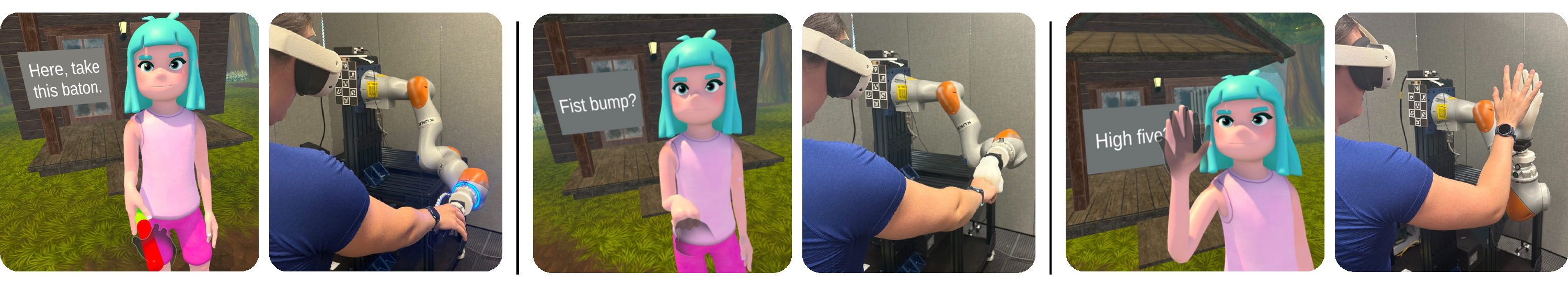}
  \caption{ETHOS (Encountered-Type Haptics for On-demand Social interaction) enables social–physical human–robot interactions in VR, enhancing both the physical and social dimensions of the user experience. Using ETHOS, we demonstrate three types of social-physical interactions with a virtual character: object handover (left), fist bump (centre), and high five (right).}
  \label{fig:ETHOS}
  \Description{Virtual and physical perspectives of three interactions with virtual characters within this work: object handover, fist bump, and high five}
\end{teaserfigure}

\renewcommand{\shortauthors}{Godden et al.}

\begin{abstract}

This work presents novel robot-mediated immersive experiences enabled by an encountered-type haptic display (ETHD) that introduces direct physical contact in virtual environments. We focus on social–physical interactions, a class of interaction associated with meaningful human outcomes in prior human–robot interaction (HRI) research. We explore the implementation of this interaction paradigm in immersive virtual environments through an object handover, fist bump, and high five with a virtual character. Extending this HRI paradigm into immersive environments enables the study of how physically grounded robotic contact and virtual augmentation jointly shape these novel social–physical interaction experiences. To support this investigation, we introduce ETHOS (Encountered-Type Haptics for On-demand Social interaction), an experimental platform integrating a torque-controlled manipulator and interchangeable props with a headset-mediated virtual experience. ETHOS enables co-located physical interaction through marker-based physical–virtual registration while concealing the robot behind the virtual environment, decoupling contact from visible robot embodiment. ETHOS supports two control strategies: static physicality (SP), providing stationary props for direct contact, and dynamic physicality (DP), which introduces motion and impact forces to approximate natural touch sensations. Technical feasibility was first characterized through spatial alignment (\(5.09 \pm 0.94\) mm) and interaction latency (\(28.58 \pm 31.21\) ms) tests, confirming average interaction performance was below discernible thresholds. To extend beyond technical feasibility, a user study then examined how robot-mediated physicality influences the developed interactions across experiential measures including presence, realism, and social connection. Participants experienced a single interaction under three conditions: a purely virtual baseline (no physicality; NP), as well as the SP and DP control strategies. Results demonstrated consistent experiential improvements from the virtual baseline to both physicality conditions, while differences between SP and DP were not conclusively observed. Overall, these findings demonstrate the feasibility and experiential value of robot-mediated social–physical interaction in VR and motivate further development of dynamic encountered-type approaches for immersive HRI.

\end{abstract}




\maketitle

\section{Introduction}
Virtual reality (VR) enables immersive interactive experiences that have been successfully applied across domains such as training \cite{howard_meta-analysis_2021}, rehabilitation \cite{howard_meta-analysis_2017}, and entertainment \cite{cimolino_you_2022}. However, these experiences remain largely dominated by the audio-visual experience, as their effectiveness often breaks down when interactions demand physical engagement \cite{wang2019multimodal}. This limitation has created a fundamental challenge for immersive experiences: many applications rely on embodied interaction, yet such interactions are difficult to render convincingly in VR. Haptic technologies (i.e., systems designed to recreate touch sensations) seek to address this gap: even simple forms of physical rendering have been shown to improve virtual task performance \cite{kreimeier_evaluation_2019} and enhance presence \cite{gibbs_comparison_2022}, the subjective sense of “being there.” Crucially, the quality of physical feedback matters: intuitive and congruent interactions enhance realism, whereas abstract or mismatched feedback can undermine immersion \cite{antona_benefits_2015}. Together, these findings position physically grounded interaction, closely resembling real-world form, as a key driver for extending the expressive and experiential capabilities of immersive experiences.

Despite this need, most existing approaches of rendering physical interaction in immersive environments have generally involved systems that provide localized or single-modality feedback, such as vibration or electrical stimulation \cite{dangxiao2019haptic}, which offer limited support for physically grounded interaction. Encountered-type haptics displays (ETHDs), in contrast, coordinate tangible props with their virtual counterparts to enable direct, co-located interaction. ETHDs typically leverage a robotic manipulator as a physical intermediary, allowing on-demand, coordinated contact that supports embodied interaction. While the visibility of the robot's physical form generally plays an important role in human-robot interaction (HRI) \cite{onnasch2021taxonomy}, ETHDs conceal themselves behind the virtual environment. This separation creates a unique opportunity to leverage the complementary strengths of HRI and immersive systems: robots provide precise, physically grounded interaction, while the virtual environment supplies rich social and contextual cues through visual and auditory stimulus that would be difficult to convey through physical embodiment alone \cite{qu2025humanoid}.

To explore this prospect, we have developed ETHOS (Encountered-Type Haptics for On-demand Social interaction), an experimental platform that integrates a torque-controlled 7-DoF robotic manipulator and interchangeable physical props with a headset-mediated VR experience to render social–physical interactions with virtual characters (i.e., object handovers, fist bumps, and high fives). Social-physical interactions (i.e., interactions with intrinsically intertwined social and physical components) represent the initial application for the developed platform, as prior HRI research has shown such interactions to elicit meaningful human outcomes, including exercise \cite{fitter2020exercising} and physical skill training \cite{block2021six}. By bridging social-physical HRI with VR for the first time, this work not only develops novel physically interactive immersive experiences, it also extends HRI experiences into new digital landscapes.

The remainder of this paper is organized as follows. Section \ref{sec:background} reviews related work in haptics technology, encountered-type haptics, and social–physical HRI. Section \ref{sec:ResearchGoals} then articulates the research goals that inform both the design and evaluation of our approach. Section \ref{sec:expPlatform} presents the ETHOS platform, followed by a technical characterization of platform performance relative to perceptual thresholds in Section \ref{sec:techEval}. Section \ref{sec:expEval} reports on a user study evaluating experiential outcomes of the developed interactions. Finally, Section \ref{sec:Discussion} interprets the findings in relation to the research goals, and Section \ref{sec:Conclusion} concludes the paper with directions for future work

\section{Background}\label{sec:background}
\subsection{Haptics in VR}
Haptics technologies are those that rendering physicality through the stimulation of our sense of touch; the human haptic system integrates signals from \textit{cutaneous} receptors in the skin and \textit{kinesthetic} receptors in muscles, tendons, and joints \cite{hale_deriving_2004}. Designing displays that address both pathways is hard in practice, so interfaces often target a subset. These focused displays inherently entail a trade-off between fidelity and versatility; feedback is either highly realistic but application-specific or abstract but broadly applicable \cite{muender2022haptic}.

Wearable and hand-held devices commonly deliver cutaneous cues in VR by stimulating the skin surface (e.g., vibrotactile motors, electrotactile arrays), which are easy to integrate and cost-effective \cite{pacchierotti2017wearable, Wearability, guo_-skin_2022}. These solutions preserve mobility but typically provide abstract sensations and limited force/proprioceptive information, which can reduce the perceived naturalness of larger-scale interactions \cite{antona_benefits_2015}. To render kinesthetic cues, grounded or body-worn kinesthetic devices apply forces and motions to the user. Commercial tabletop devices, such as Touch (3D Systems, USA) and Omega (Force Dimension, Switzerland), offer accurate force feedback over a portion of the workspace but station the user, while wearable kinesthetic systems can constrain free motion and introduce reaction forces at attachment points \cite{gu2016dexmo, choi2016wolverine}.

Multi-modal approaches combine cutaneous and kinesthetic hardware to improve realism \cite{quek2014sensory, pierce2014wearable}, but complexity and scalability become issues as the breadth of required interactions within an environment grows. This motivates the development of alternatives that have potential for greater interaction versatility while still keeping users unencumbered and providing physically meaningful contact.

\subsection{Encountered-Type Haptic Displays (ETHDs)}
ETHDs are ``\textit{systems capable of positioning all or part of themselves at an encountered location, enabling the user to voluntarily elicit haptic feedback with the environment at the appropriate time and place}'' \cite{hapticsondemand}. In practice, they co-locate physical props with virtual representations to provide passive physical interaction \cite{insko2001passive}. This is advantageous as the interaction with the prop's physical form produces the desired sensory experience and removes the need to actively apply forces or stimulation to the user. The passive nature of this feedback approach simplifies haptic design while recreating physical interaction across both cutaneous and kinesthetic sensory channels. ETHDs have been implemented using human operators \cite{cheng2015turkdeck}, drones \cite{abtahi2019beyond}, and robotic manipulators, the latter offering precise positioning and stable performance through contact. Their external grounding frees users from wearable hardware during free motion while ensuring reliable feedback at first contact. Despite these advantages, most ETHD systems remain limited to rendering static objects or surfaces \cite{mortezapoor2023cobodeck, vrrobot}. Some expand flexibility with interchangeable endpoints \cite{snakecharmer} or detachable task-relevant props \cite{hapticbutler}, but few extend towards dynamic interactions that carry significance outside the presentation of their physical form.

\subsection{Social-Physical Human-Robot Interaction}
HRI research provides valuable insights into how physicality and sociality can be combined to create richer interactive experiences. Traditionally, HRI has examined either the \textit{physical} dimension—focusing on low-level contact dynamics such as force exchange and control for tasks like object handovers \cite{chan2012grip} or collaborative manipulation \cite{ajoudani2018progress}—or the \textit{social} dimension, which investigates how behaviours such as gaze, gesture, and proxemics shape perception and trust \cite{saunderson2019robots}. Integrating these perspectives yields what we refer to as \textit{social–physical interaction}. Although the term has appeared in prior work \cite{fitter2020exercising}, many studies across assistive, instructional, and affective contexts implicitly rely on this interplay without explicitly defining it. In the absence of a formal definition, we characterize \textit{social–physical interactions} as those in which social and physical elements are intrinsically intertwined such that recreating either component in isolation fails to capture the intended experience (e.g., the sensation of knuckle impact alone does not convey the interpersonal meaning of a fist bump without its social context). 

In assistive and instructional domains, combining physical interaction with social engagement has been shown to enhance both performance and motivation. For example, Fitter et al. \cite{fitter2020exercising} used a Baxter robot to create exercise games for older adults, finding that activities blending physical contact and social cues were most engaging. Granados et al. \cite{granados2017dance} applied a similar strategy in robotic dance training, where adaptive behaviour improved skill learning. In affective domains, the physical act itself carries social meaning. Block et al. \cite{block2021six} proposed design principles for robotic hugging that combine physical qualities (e.g., softness, warmth) with social adaptation (e.g., adjusting to user size) to improve experience. Okamura et al. \cite{okamura2017design} demonstrated how high-five gestures, when implemented in robots, embody both physical and social dimensions, enhancing engagement and communicative value.

The demonstrated benefits of social–physical interaction in HRI motivate extending these interaction paradigms into immersive virtual environments, where realistic physical rendering remains a fundamental challenge. ETHDs provide a natural bridge for this translation by enabling physical HRI within an immersive context. Recreating social–physical interactions through an ETHD can enable physically interactive immersive experiences while also introducing novel HRI paradigms in which the robot is concealed from the user. This integration creates opportunities to investigate how physical and virtual elements can be designed and coordinated to shape user experience and application outcomes. To the authors’ knowledge, this work represents the first integration of encountered-type haptics and social–physical interaction, exploring how their complementary strengths can be combined to enable engaging and embodied interaction experiences.

\section{Research Goals and Contributions}\label{sec:ResearchGoals}
To enable novel, embodied immersive experiences, we argue that an ETHD capable of supporting social–physical interactions in VR is a promising direction. Such interactions are inherently dynamic and multifaceted, and cannot be adequately rendered by existing ETHDs, which primarily focus on the stationary presentation of simple objects. This limitation motivates the development of ETHOS, an experimental platform designed to explore how social–physical interactions can be translated into VR through physical HRI.

Accordingly, we formulate the following research questions (RQs) to guide our investigation.

\begin{enumerate}[label=\textbf{RQ\arabic*:}, leftmargin=*]
    \item What design considerations are required for an encountered-type haptic display to support dynamic social–physical interactions in immersive virtual reality?
    \item What levels of spatial and temporal accuracy can such a system achieve?
    \item How are social–physical interactions experienced within an immersive virtual environment?
    \item How do different levels of physical rendering provided by an ETHD influence user experience during virtual social–physical interaction, as measured by related experiential metrics (e.g., presence, enjoyment)?
\end{enumerate}


\subsection{Contributions}
This work makes the following contributions:
\begin{enumerate}[leftmargin=*]
    \item The design and implementation of ETHOS, an ETHD capable of supporting dynamic social–physical interactions in immersive virtual reality.
    \item A technical characterization of ETHOS, quantifying its spatial and temporal accuracy relative to perceptual thresholds.
    \item An empirical evaluation of social–physical interactions in immersive VR, examining how different levels of physical rendering influence user experience.
\end{enumerate}

\section{Experimental Platform Design}\label{sec:expPlatform}
ETHOS (Encountered-Type Haptics for On-demand Social interaction) is an experimental platform designed to support interactive immersive experiences that are both physically precise and socially meaningful. The platform integrates a physical subsystem—a robotic manipulator outfitted with customized physical props—and a virtual subsystem—a headset-mediated immersive experience presenting the environment and virtual characters—linked through a fiducial-based registration process and UDP communication enabling bidirectional exchange of subsystem state information. In this work, ETHOS is applied to three social–physical interactions with a virtual character: object handover, fist bump, and high five.

\begin{figure}[t]
    \centering
    \includegraphics[width=0.9\linewidth]{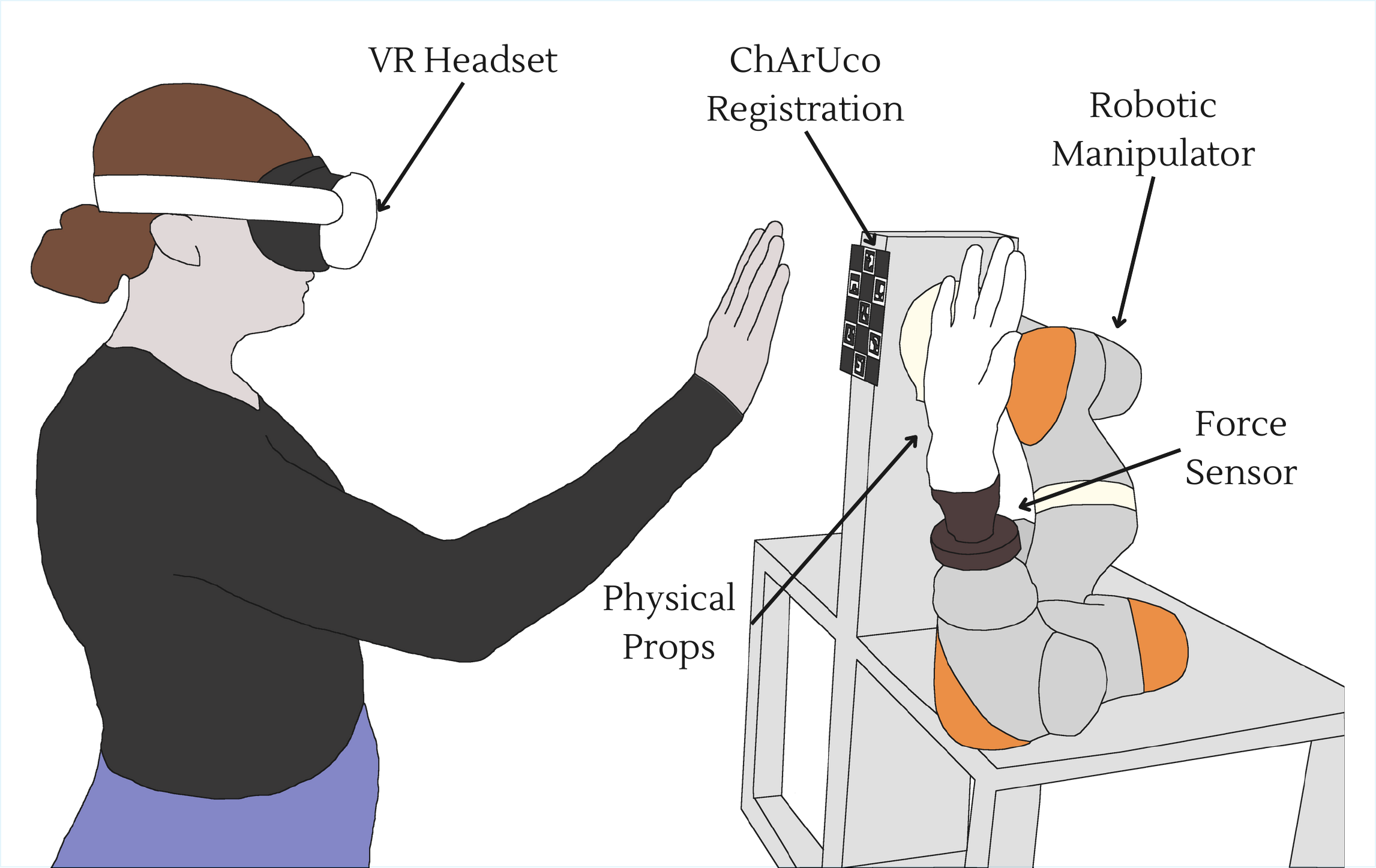}
    \caption{Overview of the ETHOS system design. The setup integrates a VR headset for immersive visualization and user hand tracking, ChArUco-based registration to establish a shared anchor between the physical and virtual environments, a robotic manipulator equipped with interchangeable physical props to render on-demand physical contact, and a force sensor to monitor interaction forces.}
    \label{fig:system}
\end{figure}

By leveraging an anthropomorphic virtual character, the interaction experience provides recognizable non-verbal social cues that mitigate challenges often associated with non-anthropomorphic robots \cite{qu2025humanoid}, while still delivering tangible physical contact through encountered-type interactions. Unlike typical HRI systems that emphasize transparency through robot awareness and shared intent \cite{lyons2014transparency}, ETHOS deliberately conceals the robot behind the virtual experience, directing user attention toward the avatar rather than the machine. This hidden embodiment places increased demands on precise synchronization between physical and virtual domains, while creating opportunities to study how their coordinated design can shape user experience and application outcomes. Figure~\ref{fig:ETHOS} illustrates ETHOS in action, while Figure~\ref{fig:system} summarizes the platform’s principal components across both subsystems.


\subsection{Physical Subsystem}\label{subsec:PhySubsys}
The physical subsystem, responsible for rendering tangible contact, consists of interchangeable props mounted on the end effector of a torque-controlled LBR iiwa 7 R800 robotic manipulator (KUKA, Germany). The robot is mounted horizontally on a pedestal and controlled in real time at 1000~Hz via the Fast Robot Interface (FRI) \cite{schreiber2010fast}, connected to a desktop workstation (Intel Core i7-14700, 4.2~GHz, 32~GB RAM) running Ubuntu~22.04 with a real-time kernel and integrated with ROS~2 Humble. A six-axis Mini40 force–torque sensor (ATI Industrial Automation, USA) with a quick-release mounted at the base of the robot's end-effector enables secure prop attachment and rapid swapping, with internal compensation applied to ensure accurate contact force measurements.  

\begin{figure}[t]
    \centering
    \includegraphics[width=0.9\linewidth]{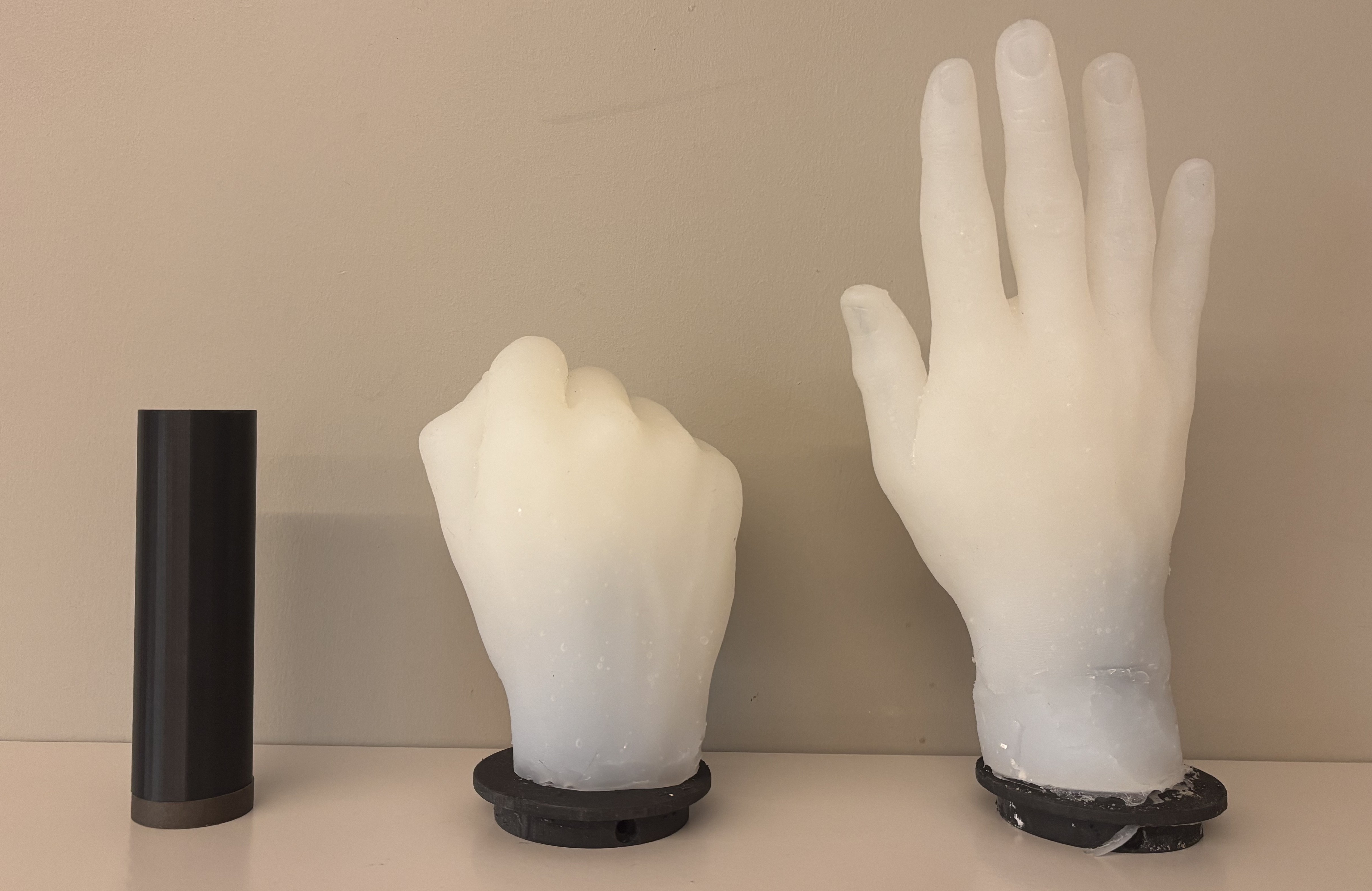}
    \caption{Physical props developed for ETHOS — a baton (left), silicone fist (centre), and silicone open hand (right)—created for the handover, fist bump, and high five interactions, respectively.}
    \label{fig:props}
\end{figure}

To match the selected interactions, three props have been developed as shown in Figure \ref{fig:props}: a 3D-printed baton for handovers, a silicone fist for fist bumps, and a silicone open hand for high fives. Both hand models are cast using Dragon Skin FX-Pro\footnote{https://www.smooth-on.com/products/dragon-skin-fx-pro/} in their respective configurations and reinforced with galvanized steel wire to emulate internal bone structure.

\subsection{Virtual Subsystem}
The virtual subsystem, responsible for presenting the audio-visual experience to the user, uses a Meta Quest~3 head-mounted display (HMD), selected for its accessibility, popularity, and native markerless hand-tracking that has been characterized in prior work \cite{godden2025robotic}. The virtual environment is developed in Unity~6.1 (6000.1.1f1) on a Windows~11 x64 laptop (AMD Ryzen 5800H with Radeon Graphics, 3.2~GHz, 16~GB RAM, NVIDIA GeForce RTX~3060 Laptop GPU). A Unity project containing the virtual scene is built as an Android application and deployed directly to the HMD. Camera rig control, passthrough functionality, and inside-out hand tracking are provided by the Meta XR All-in-One SDK (76.0.1)\footnote{\url{https://developers.meta.com/horizon/downloads/package/meta-xr-sdk-all-in-one-upm/76.0.1}}, with deployment managed through the OpenXR plugin\footnote{\url{https://docs.unity3d.com/Packages/com.unity.xr.openxr@1.14/manual/index.html}} and XR Plugin Management Package (4.5.1)\footnote{\url{https://docs.unity3d.com/Packages/com.unity.xr.management@4.5/manual/index.html}}. Meta-specific functionality is enabled through the Meta XR Feature Group.  


The virtual scene, developed in Unity and delivered through the HMD, places the user within a forest environment. This setting was selected as a visually neutral context with naturally occurring ambient sounds (e.g., bird chirping), which were used to mask mechanical noise generated by the robot during interaction. At the start of the experience, users were free to explore the scene visually before walking to a predefined interaction location indicated by a coloured circle on the ground. Once properly positioned, a virtual character entered the scene by exiting a nearby cabin, preserving the plausibility of the character’s arrival.

From this point onward, the virtual character led the interaction through a combination of behavioral and text-based cues. Since the robot remained hidden from the user’s view, all social cues were conveyed exclusively through the virtual character’s behaviour. Three cues were implemented: (1) a greeting gesture, consisting of a wave accompanied by eye contact upon entry; (2) a readiness cue, in which the character raised its arm into an interaction-appropriate pose (e.g., an outstretched arm holding a baton for handover \cite{strabala2013toward}); and (3) a gaze pattern that alternated between the interaction site and the user to reinforce the intended interaction and readiness of the character, following the approach of Moon et al. \cite{moon2014meet}. Together, these cues provide socially recognizable signals intended to support the naturalness and believability of social–physical interactions with a virtual character. More broadly, they serve as an initial exploration of how virtual social cues can be coordinated with physical interaction to shape the overall user experience.

\subsection{Physical-Virtual Coordination Subsystem}\label{subsec:PVCoordination}
Accurate spatial registration is required to co-locate the virtual and physical environments. Instead of relying on resource-intensive systems like motion capture \cite{pan2017catching} or external depth sensors \cite{7563561}, we implement a lightweight fiducial-based method using a ChArUco board\footnote{\url{https://docs.opencv.org/3.4/df/d4a/tutorial_charuco_detection.html}} and the HMD’s onboard cameras. ChArUco detection and pose estimation are performed using Quest ArUco Marker Tracking\footnote{https://github.com/TakashiYoshinaga/QuestArUcoMarkerTracking} with OpenCV for Unity\footnote{https://enoxsoftware.com/opencvforunity/}. A single registration procedure establishes a reference transform based on the ChArUco board, serving as a shared anchor between the virtual and physical domains.  

Headset tracking data and synchronization signals are transmitted to the robot at the HMD’s $\sim$90~Hz refresh rate, while end-effector position is streamed back at the $\sim$1000~Hz robot control cycle. To ensure robust communication, key messages are redundantly transmitted and serialized using Google’s Protocol Buffer Framework\footnote{https://github.com/protocolbuffers/protobuf} over connectionless UDP unicast. This framework provides reliable colocation and temporal sequencing between the virtual and physical subsystems. Safety was enforced through shared message passing, gating robotic motion based on user positioning captured by the headset's native tracking. During an interaction, the user is first positioned at a location outside the robot's workspace, but still within arm's reach of the eventual interaction location, before robotic motion and interaction sequencing begin. User positioning is then continuously monitored through interaction to halt motion if the user's body/head enters the robot's workspace (minimizing the risk of the most severe collisions). An external operator is also monitoring the experience with an emergency stop to provide an additional safeguarding layer.

\subsection{Interaction Control Subsystem}\label{subsec:control}
Our system implements two strategies for robotic control in the phases before and during interaction: static physicality (SP) and dynamic physicality (DP). In both strategies, the system continuously monitors the force at the end effector to detect user contact. Thresholds for natural-feeling contact were determined through informal pilot testing by members of the development team, who iteratively evaluated and fine-tuned force limits. Based on this process, thresholds were set to 7.5~N for object handovers and 15.0~N for both fist bumps and high fives, reflecting the greater impact of the latter gestures. Exceeding these thresholds signals completion of the interaction.


\subsubsection{Static Physicality (SP)}
The SP strategy aligns with the state-of-the-art for ETHDs by presenting a stationary object for the user to interact with. By raising its arm after approach, the virtual avatar's arm aligns with the physical prop, enabling the robotic manipulator to remain stationary throughout the physical interaction.

\subsubsection{Dynamic Physicality (DP)}
\begin{figure*}[t]
  \centering
  \subfloat[]{%
    \includegraphics[angle=-90,width=0.45\textwidth]{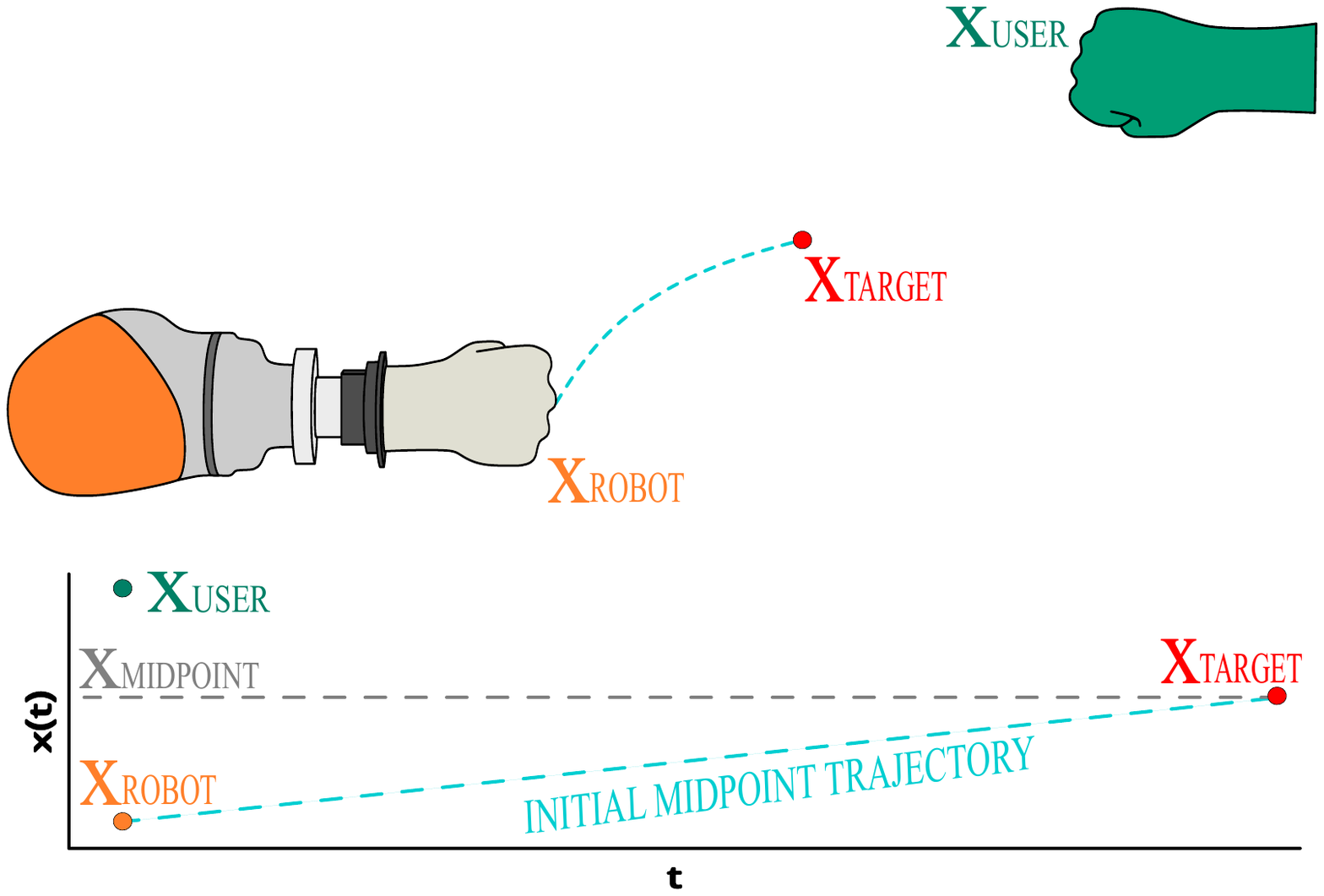}%
  }\hspace{0.03\textwidth}
  \subfloat[]{%
    \includegraphics[angle=-90,width=0.45\textwidth]{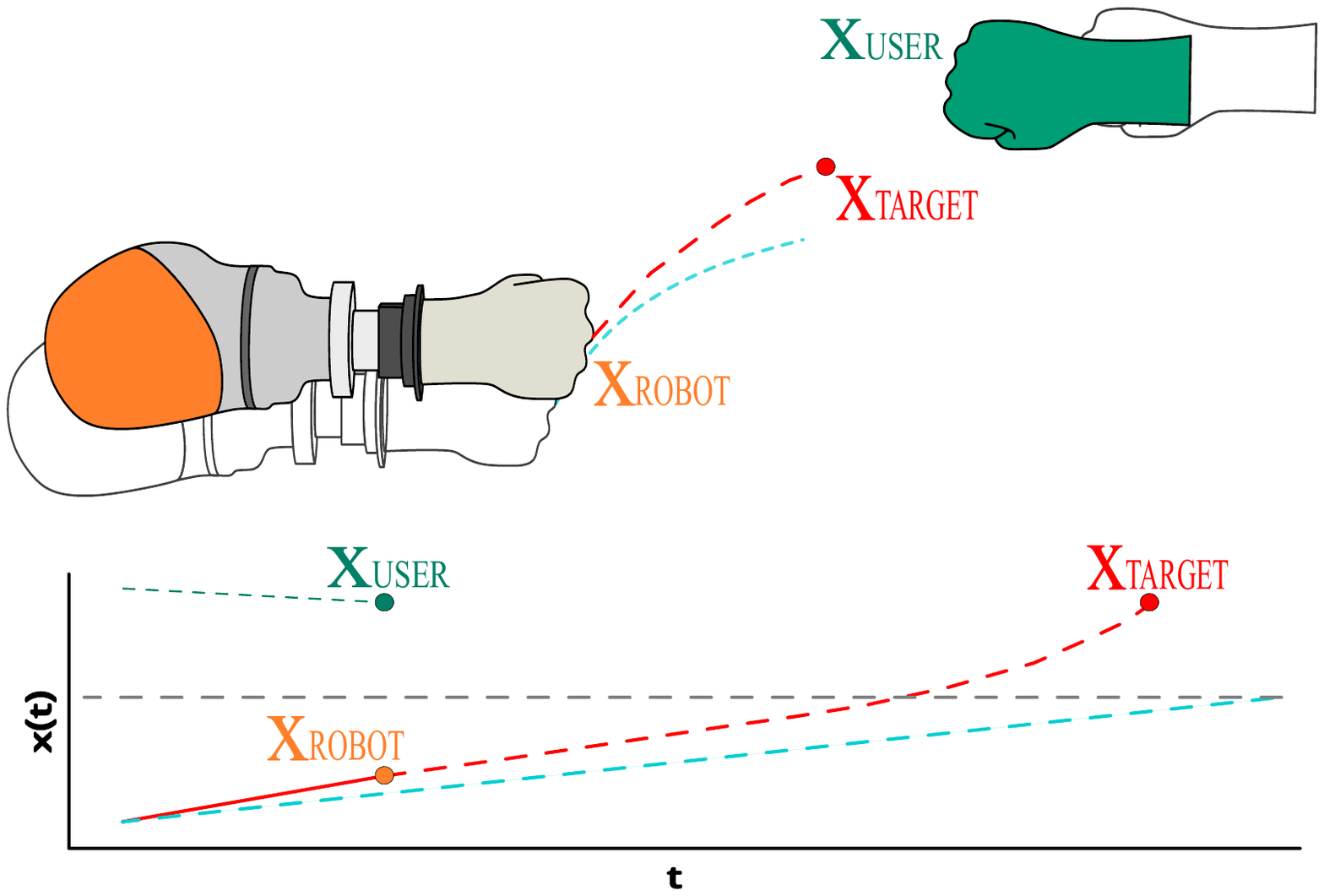}%
  }\\[1ex]

  \subfloat[]{%
    \includegraphics[angle=-90,width=0.45\textwidth]{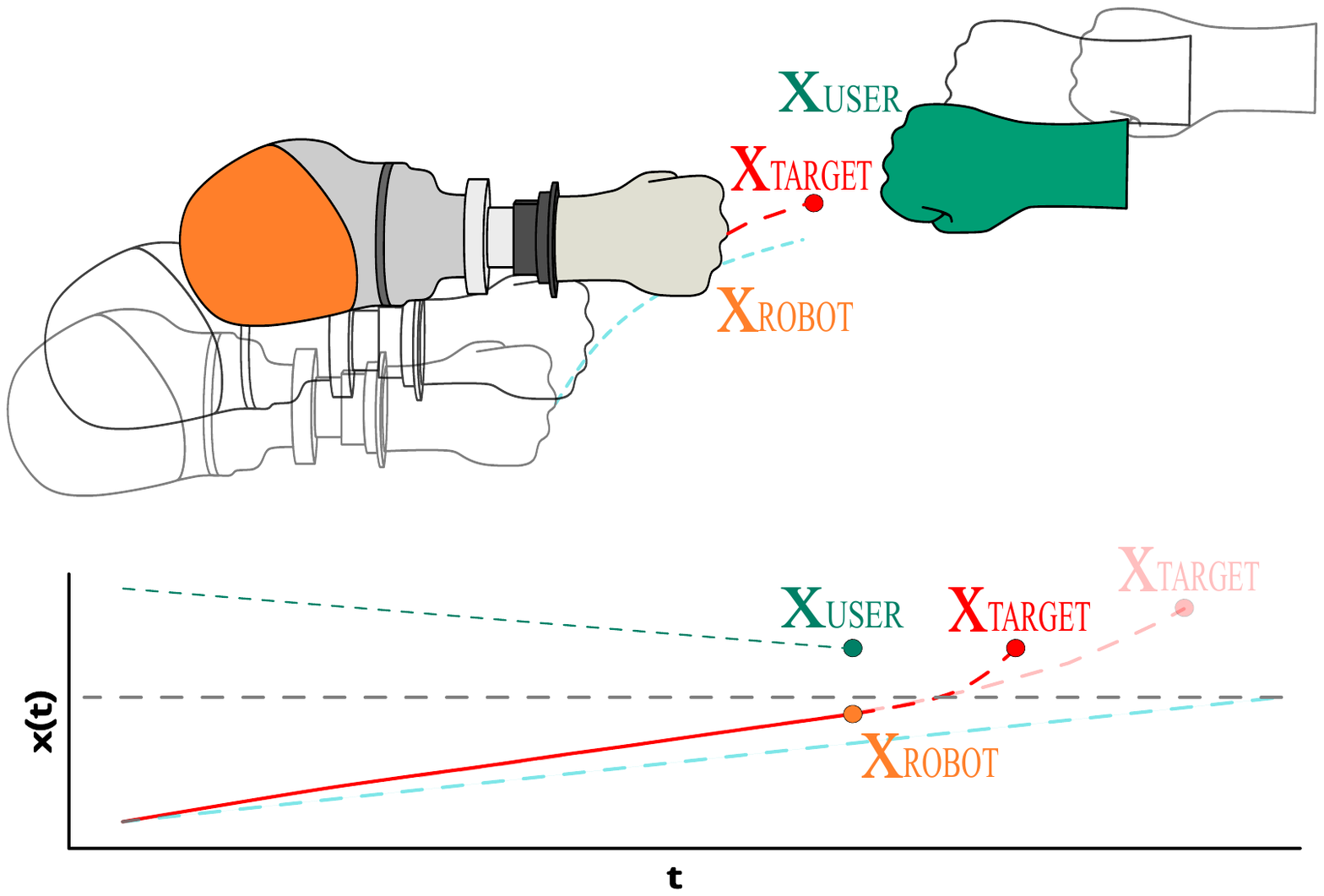}%
  }\hspace{0.03\textwidth}
  \subfloat[]{%
    \includegraphics[angle=-90,width=0.45\textwidth]{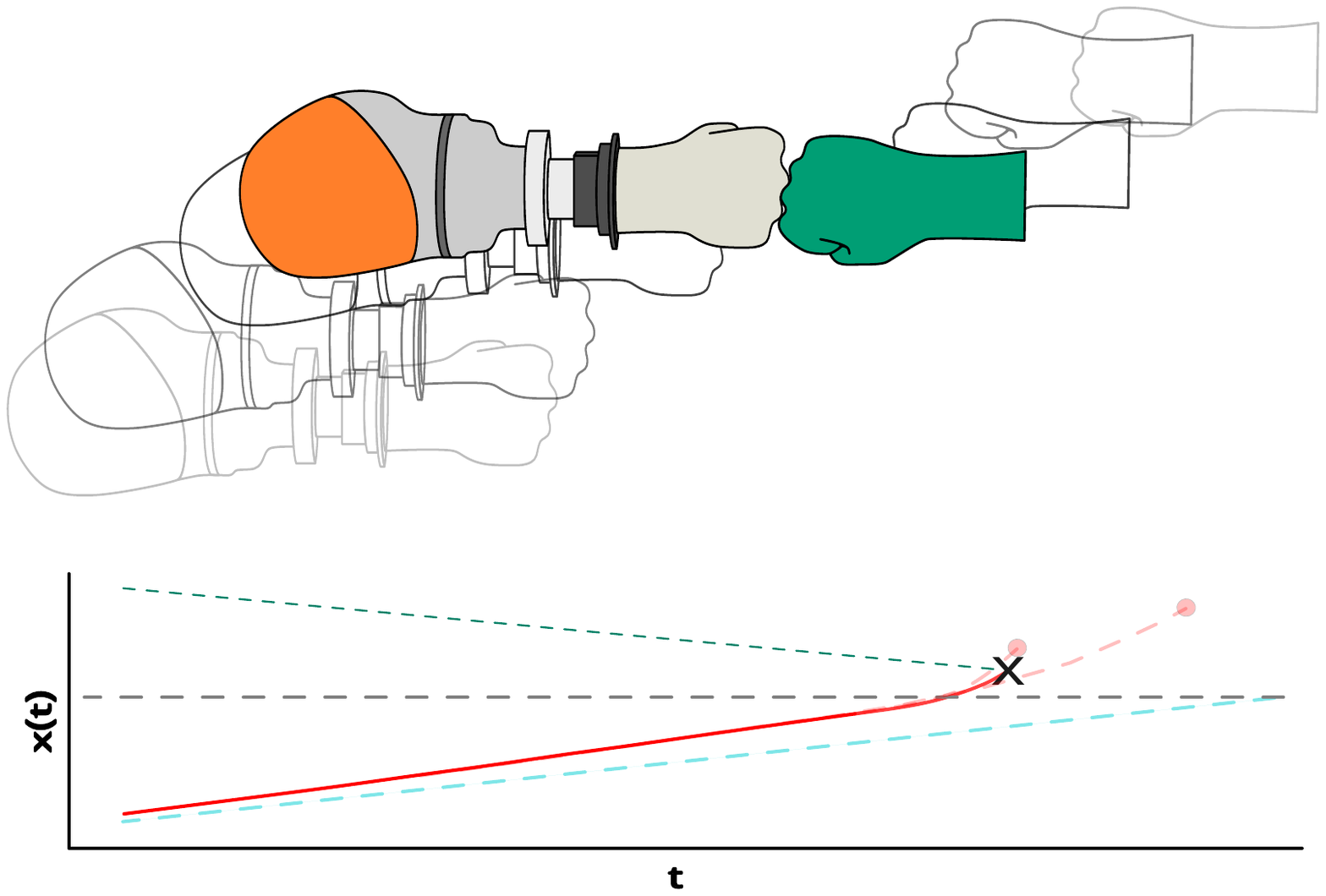}%
  }

  \caption{Visualization of the dynamic trajectory generation strategy across interaction phases. Each panel illustrates the physical interaction state (top) alongside a corresponding single-axis trajectory as a function of time (bottom). (a) When the user’s hand first approaches, an initial midpoint trajectory is generated between the robot and the target. (b) As the interaction progresses, the midpoint trajectory is exponentially blended toward the user’s current hand position. (c) Continued blending adapts the trajectory in response to ongoing user motion. (d) The interaction concludes when physical contact is reached at a unique, user-dependent location.}
  \label{fig:dynamicTraj}
\end{figure*}

The DP strategy, in contrast, leverages online hand pose tracking to create a unique contact point for each interaction. This strategy was inspired by the trajectory generation of Pan et al. \cite{FastHandovers}, with an adjustment to the update of the weight term throughout interaction. At the start of the interaction phase, the prop is positioned at the same pre-defined location as in the static strategy. When the user’s hand enters the 3-dimensional interaction volume, defined as a 10 × 30 cm rectangle extending 30 cm forward from the prop's coordinate frame, the robot computes an initial one-second trajectory that targets the midpoint between the prop and the hand at that instant. The target location is then continuously updated as a weighted average of the initial trajectory and the user’s current hand position, as shown in Equation \ref{eq:dynpos}. The weight term is defined in Equation \ref{eq:weight}, where $t_{dynamic}$ denotes the time elapsed since dynamic tracking began. This formulation applies an exponential weighting, prioritizing smooth early motion by favouring the initial trajectory and gradually shifting toward the user’s actual hand position as the interaction progresses. The weight is capped at a maximum value of 1, ensuring that if the interaction extends beyond the one-second allocation, the system continues to track the user’s current hand position. A visualization of the progression of this tracking is provided in Figure \ref{fig:dynamicTraj}.

\begin{equation} \label{eq:weight}
w(t) = \frac{e^{3 \alpha(t)} - 1}{e^{3} - 1}, \quad 
\alpha(t) = \min\!\bigl(t_{\text{dynamic}},\,1\bigr)
\end{equation}

\begin{equation} \label{eq:dynpos}
    x_{target}(t) = (1 - w(t))x_{midTraj}(t) + w(t)x_{hand}(t)
\end{equation}


\section{Technical Evaluation}\label{sec:techEval}
The first step in evaluating ETHOS's performance was to define technical feasibility tests to determine both spatial and temporal alignment of the interactions' physical and virtual presentations. These feasibility metrics enabled performance benchmarking against perceptual thresholds to evaluate whether the inevitable error introduced by the platform is acceptable. In development of these tests, we looked to the work of Nilsson et al. who proposed two criteria for success when using physical props for rendering virtual interaction: similarity and colocation \cite{nilsson_propping_2021}. The criterion of similarity refers to the physical object's material and geometric properties matching those of its virtual counterparts, whereas colocation refers to the spatial alignment of virtual and physical forms to allow seamless interaction. The focus of our initial system characterization presented here is colocation; without proper colocation, the interactions may not be feasible at all, leaving the props' physical properties secondary. Although similarity is not directly evaluated here, the props have shown promising initial success in recreating convincing interactions, suggesting their suitability for the present study while warranting further testing in future work.

\subsection{Spatial Alignment}
Following Nilsson et al., we evaluated the spatial alignment of the virtual and real-world environments/props under static conditions \cite{nilsson_propping_2021} as a basic test of co-location. This evaluation deliberately excludes temporal factors (e.g., message-passing latency) in order to isolate the performance of our fiducial-based virtual–physical registration. Given the high positioning accuracy of the robotic manipulator and the fact that the marker’s relative placement was established using motion capture, the dominant source of error in static colocation arises from the registration process itself. Since this process depends on computer-vision tracking of the ChArUco board, the estimated anchor point between the physical and virtual environments can vary across trials, motivating a dedicated evaluation of registration accuracy.

To provide ground truth for our measurements, a retroreflective marker-based motion-capture system, Vicon Vero (v2.2)\footnote{https://www.vicon.com/hardware/cameras/vero/}, was used. Vicon Tracker Software (3.9.0.128020h)\footnote{https://help.vicon.com/space/Tracker39} was run on a Windows 11 x64 system (Intel Xeon W-2123 3.6GHz, 16GB RAM) to provide sub-millimetre error tracking of marker constellations aligned with the ChAruCo board and the VR controller. 

\begin{figure}[t]
    \centering
    \includegraphics[width=0.9\linewidth]{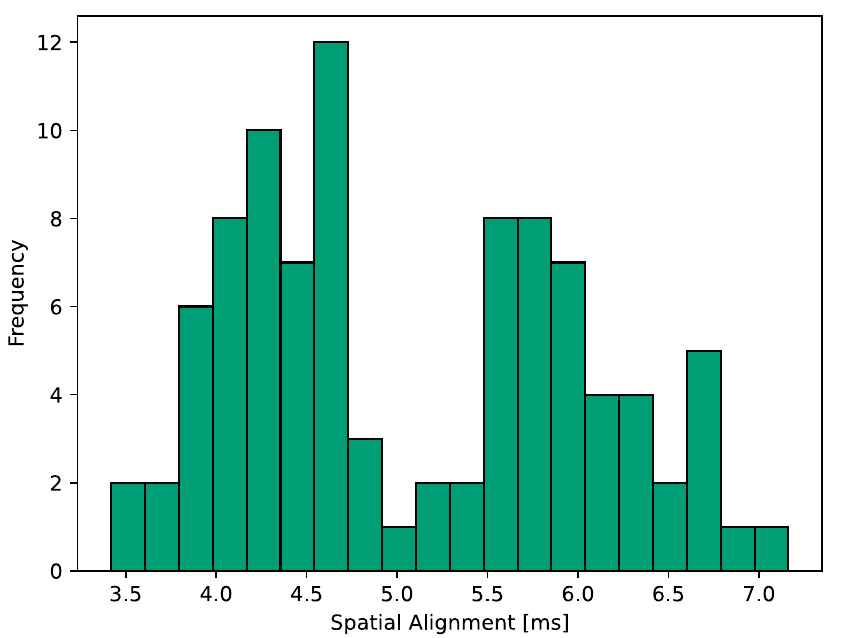}
    \caption{Histogram of static spatial alignment error between physical and virtual objects after fiducial-based registration. Error values reflect Euclidean distance between corresponding measurements across 20 registration trials.}
    \label{fig:spatialAlignmentHist}
\end{figure}

With markers placed on the ChAruCo board and VR controller, both physical and virtual positions of these objects could be tracked. The spatial alignment could then be determined by registering and recording the error between each measurement frame's relative positions with respect to the objects. This process was completed across 20 registration trials, recording positional error 5 times per trial (100 total measurements). Over the registration trials, the \textit{spatial alignment} of the physical-virtual registration was determined to be $5.09 \pm 0.94 mm $, visualized in Figure \ref{fig:spatialAlignmentHist}.

\subsection{Interaction Latency}
With static spatial alignment characterized, temporal factors must also be considered when evaluating colocation. Interaction latency—defined as the time difference between a physical event and its corresponding virtual representation—can significantly influence the perceived realism of haptic interaction. In our system, latency arises from multiple sources, including message-passing delays between subsystems, processing overhead within the control and rendering pipelines, and the inherent delay introduced by markerless hand tracking \cite{godden2025robotic}. While a component-wise analysis of these sources would be informative, characterizing end-to-end latency is of primary importance for this initial system evaluation, as it reflects the cumulative delay the user ultimately experiences.

Interaction latency was measured by comparing the timing of hand–object collisions in the physical and virtual environments. A physical collision was defined as the moment when the measured interaction force exceeded the corresponding interaction threshold, while a virtual collision was identified by the intersection of the user’s hand with the object’s mesh. The collision signals were coordinated with the message-passing framework described above in Section \ref{subsec:PVCoordination} and the time difference between signals was recorded as the interaction latency of the trial. Four pilot participants completed 25 trials under static physicality and 25 trials under dynamic physicality for each interaction type, yielding a total of 600 latency measurements. The \textit{temporal alignment}, reported as the mean ± 1 standard deviation, for each interaction type under each control strategy is shown in Table \ref{tab:latency}.

\begin{table}[htbp]
\caption{Mean interaction latency results with one standard deviation for handover, fist bump, and high five across static and dynamic control conditions}
\begin{center}
\resizebox{\columnwidth}{!}{
\begin{tabular}{|c|c|c|}
\hline
\textbf{Interaction Type}&\textbf{Static Latency (ms)}&\textbf{Dynamic Latency (ms)} \\
\hline
Handover   &  18.83 $\pm$ 40.65 &  25.72 $\pm$ 35.48 \\
\hline
Fist Bump &  36.01 $\pm$ 35.45 &  24.49 $\pm$ 18.77 \\
\hline
High Five  &  23.29 $\pm$ 14.99 &  43.11 $\pm$ 27.02\\
\hline
\end{tabular}
}
\label{tab:latency}
\end{center}
\end{table}

\section{Experiential Evaluation}\label{sec:expEval}
To move beyond technical feasibility, we conducted a user study to empirically investigate user experience within the developed social-physical interactive experience and how varying levels of physicality rendered through ETHOS influences them. The following subsections describe the experimental conditions, evaluation metrics, participant demographics, study procedure, and corresponding results.
                                                                
\subsection{Experimental Conditions}
Within the experiential evaluation, \textit{interaction type} was treated as an experimental factor with three categories: object handover, fist bump, and high five. These interactions were selected as they are brief, socially recognizable, and well suited to an ETHD approach. More complex interactions, such as hugging or handshaking, require nuanced force modulation and physical adaptation that remain challenging for robotic systems to reproduce convincingly \cite{block2021six, knoop2017handshake}. Evaluating \textit{interaction type} therefore allows us to assess both the suitability of the proposed approach for rendering interpersonal gestures and whether interaction-specific characteristics (e.g., contact form and location) meaningfully influence user experience, or whether the system generalizes across interaction classes.

Each interaction was evaluated under three \textit{physicality conditions}, which varied the behaviour of the robot in producing physical contact:

\begin{enumerate}
    \item \textit{No Physicality (NP)} interactions were purely virtual, providing no haptic feedback and serving as a baseline for interaction with a virtual character.
    \item \textit{Static Physicality (SP)} introduced stationary alignment of a physical prop with its virtual counterpart, following the static control strategy described in Section~\ref{subsec:control}.
    \item \textit{Dynamic Physicality (DP)} extended the static condition by incorporating motion and impact behaviours commonly observed in human–human interaction \cite{strabala2013toward}. Following the dynamic control strategy outlined in Section~\ref{subsec:control}, this condition produces an interaction-specific contact point and more closely approximates the natural form of these gestures.
\end{enumerate}

With three interaction types and three physicality conditions, we employed a mixed factorial study design. Each participant experienced a single interaction type (between-subjects) across all three physicality conditions (within-subjects). This design enabled evaluation of the main effects of \textit{interaction type} and \textit{physicality condition}, as well as their interaction effects.

\subsection{Evaluation Metrics}
To assess experiential outcomes, we first measured \textit{presence}, a central construct in virtual reality and mediated interaction research. Presence broadly refers to the psychological perception of “being there” within a mediated environment. Among the many existing definitions, we adopt that of Lee \cite{lee2004presence}, who defines presence as \textit{a psychological state in which virtual objects are experienced as actual objects in either sensory or nonsensory ways}. Lee further distinguishes between physical, social, and self-presence.

Building on this framework, Makransky et al.~\cite{makransky2017development} developed the validated 15-item Multimodal Presence Scale (MPS), which captures these sub-dimensions and their associated attributes (Table~\ref{tab:MPS}). The MPS was selected due to its explicit differentiation between physical, social, and self-presence, making it well-suited for evaluating social–physical interactions in immersive environments.

\begin{table}[ht]
\centering
\caption{Sub-dimensions and associated area attributes of the Multimodal Presence Scale (MPS) by Makransky et al.~\cite{makransky2017development}.}
\resizebox{\columnwidth}{!}{
\begin{tabular}{@{}ll@{}}
\toprule
\textbf{Sub-dimension} & \textbf{Area attribute} \\
\midrule
\multirow{5}{*}{Physical presence} &
  Physical realism (PR) \\
& Not paying attention to real environment (NARE) \\
& Control/act in the virtual environment (CA) \\
& Sense of being in the virtual environment (SBVE) \\
& Not aware of the physical mediation (NAPM) \\
\midrule
\multirow{4}{*}{Social presence} &
  Sense of coexistence (CE) \\
& Human realism (HR) \\
& Not aware of the artificiality of social interaction (NAASI) \\
& Not aware of the social mediation (NASM) \\
\midrule
\multirow{4}{*}{Self-presence} &
  Sense of bodily connectivity (SBC) \\
& Sense of bodily extension (SBE) \\
& Emotional connectivity (EC) \\
& Sense of self being in the virtual environment (SSBVE) \\
\bottomrule
\end{tabular}
}
\label{tab:MPS}
\end{table}

To complement and more directly assess the social–physical interactions, additional custom items were included to capture perceived realism, enjoyment, comfort, and connection to the virtual character. These items were phrased as: \textit{The interaction felt realistic}, \textit{The interaction was enjoyable}, \textit{I felt comfortable during the interaction}, and \textit{My connection to the character was strengthened by the interaction}. All items were rated on a 7-point Likert scale ranging from strongly disagree (1) to strongly agree (7).

\subsection{Participants}
This study was reviewed and approved by the Queen's University General Research Ethics Board. An a priori power analysis was conducted to determine the required sample size. With $\alpha = 0.05$, a minimum of 54 participants (18 per interaction type) was required to detect a moderate effect size ($f = 0.25$) with $90\%$ power ($1 - \beta$), assuming a correlation among repeated measures of 0.5 and a nonsphericity correction of $\epsilon = 0.75$.

Participants were recruited from within Queen's University. In total, 55 participants (24 female, 31 male), aged 18–53 ($M = 26.84$, $SD = 7.97$), took part in the study. Data from one participant were excluded due to technical difficulties during their session. All participants provided informed consent and were informed that participation was voluntary and that they could withdraw at any time without penalty. The presence of the robot and the possibility of physical contact with its end effector in certain conditions were disclosed prior to participation. Participants were also informed about the use of noise-cancelling headphones. Although concealing the robot and masking its operational sounds enhanced immersion, these measures reduced transparency to the physical environment, a trade-off that was explicitly discussed with participants. Permission was obtained to collect video recordings and direct quotations from experimental sessions. No incentives or compensation were provided.

\subsection{Procedure}
Each session began with an overview of the study and a walkthrough of the interaction procedure. Participants reviewed and signed an informed consent form and completed a pre-study demographic questionnaire. They were then fitted with the VR headset and noise-cancelling headphones, which provided ambient audio (e.g., forest sounds) while masking robot-generated noise. Participants were given time to acclimate to the virtual environment and the markerless hand-tracking system. Following acclimation, physical–virtual registration was performed as described in Section~\ref{subsec:PVCoordination}.

Participants subsequently completed three trials of their assigned interaction type, one under each physicality condition. The order of physicality conditions was counterbalanced using a Latin square design to mitigate order effects. For each trial, participants were guided to a safe interaction location outside the robot’s workspace while remaining within reach of the interaction site. The virtual character then entered the scene and initiated the interaction. Upon completion of each trial, participants completed the questionnaire within the VR environment, consistent with prior findings that in-situ administration yields more reliable presence measures \cite{schwind2019using}. After completing all trials, participants provided open-ended feedback. Each session lasted approximately 30 minutes.

\subsection{Results}
\begin{figure*}[t]
  \centering
  \subfloat[Total]{%
    \includegraphics[width=0.40\textwidth]{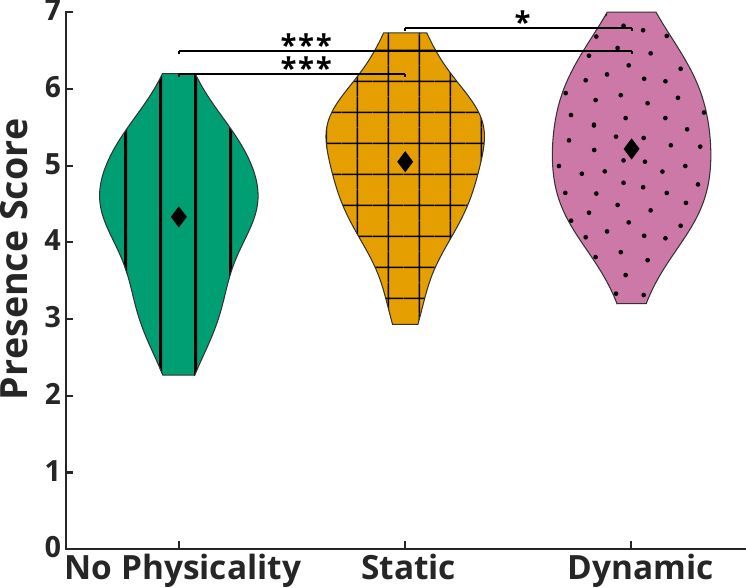}%
  }\hspace{0.10\textwidth}
  \subfloat[Physical]{%
    \includegraphics[width=0.40\textwidth]{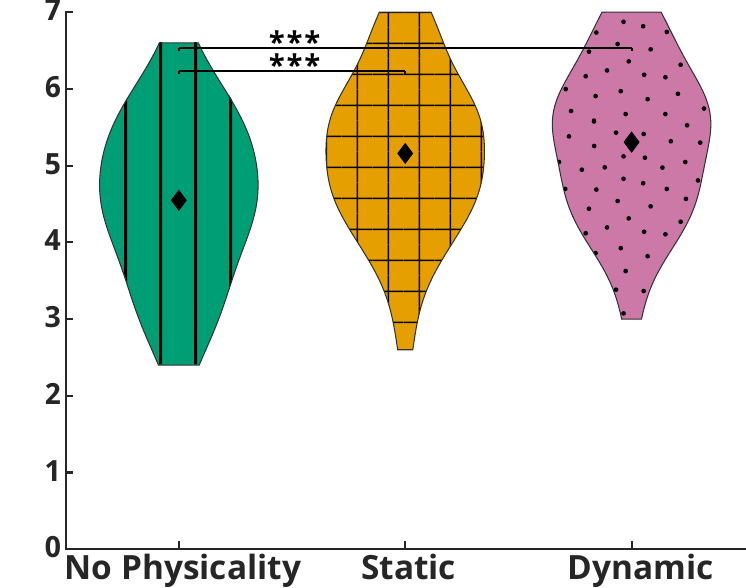}%
  }\\[1ex]

  \subfloat[Social]{%
    \includegraphics[width=0.40\textwidth]{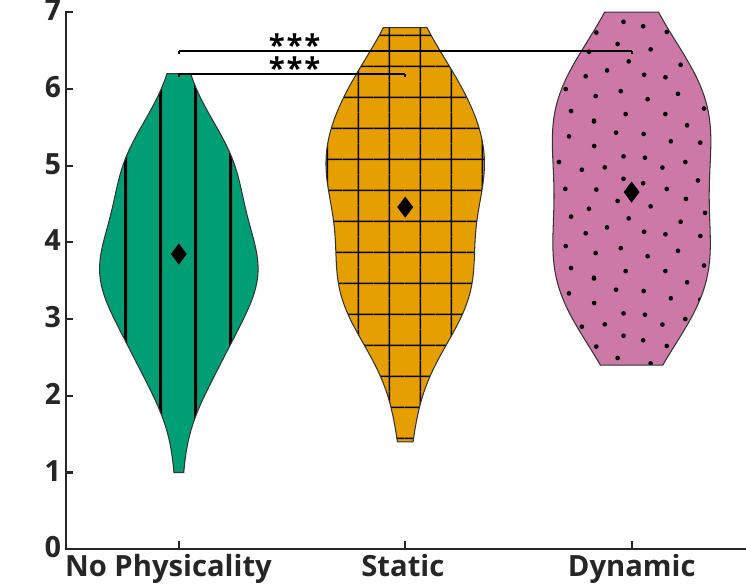}%
  }\hspace{0.10\textwidth}
  \subfloat[Self]{%
    \includegraphics[width=0.40\textwidth]{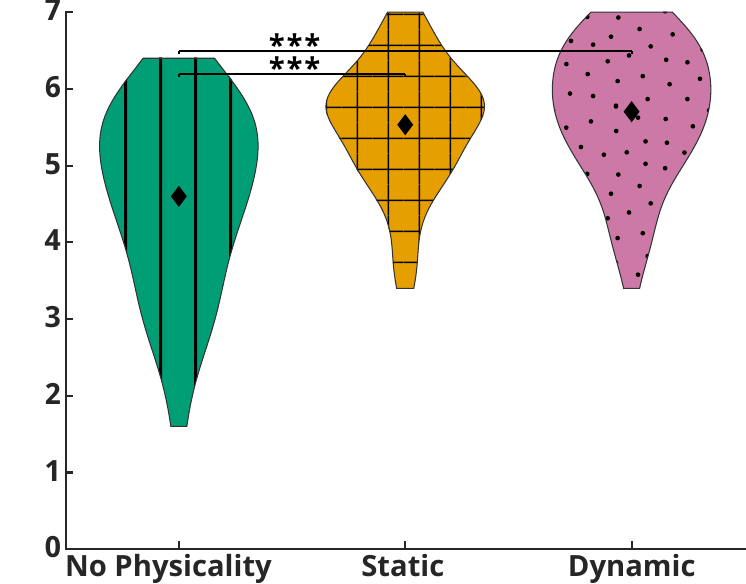}%
  }

  \caption{Presence results visualized across physicality conditions—no physicality, static physicality, and dynamic physicality—for the Multimodal Presence Scale (MPS) \cite{makransky2017development}: (a) total presence, (b) physical presence, (c) social presence, and (d) self-presence. Significant between-condition differences are annotated (* $p<0.05$, *** $p<0.001$).}
  \label{fig:presence}
\end{figure*}

\begin{figure}[t]
    \centering
    \includegraphics[width=0.9\linewidth]{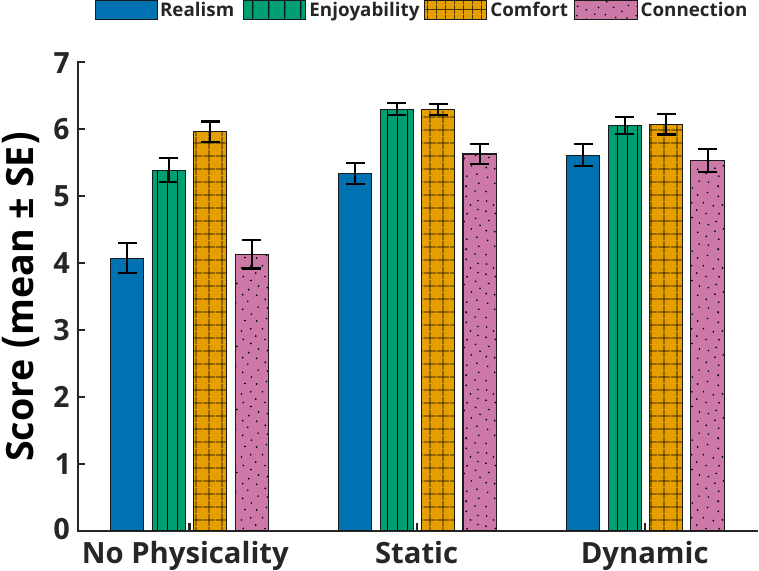}
    \caption{Mean ratings of realism, enjoyability, comfort, and character connection across the three physicality conditions (no physicality, static, dynamic). Error bars represent ± standard error of the mean (SEM).}
    \label{fig:secondaryMetrics}
\end{figure}

Given the mixed factorial design of the study, all effects were analyzed using mixed ANOVAs. This approach enabled assessment of the within-subject effect of \textit{physicality condition}, the between-subject effect of \textit{interaction type}, and their interaction. Mauchly’s test of sphericity was violated in all cases ($p < 0.05$); accordingly, Greenhouse–Geisser corrections were applied throughout. Unless otherwise stated, all reported results reflect these corrections. Effect sizes are reported as partial eta squared ($\eta_p^2$), with values of 0.01, 0.06, and 0.14 corresponding to small, medium, and large effects, respectively \cite{cohen2013statistical}. Presence results across physicality conditions are visualized in Figure~\ref{fig:presence}, while realism, enjoyability, comfort, and character connection are shown in Figure~\ref{fig:secondaryMetrics}.

\subsubsection{Presence}
As the Multimodal Presence Scale (MPS) \cite{makransky2017development} provides both an overall presence score and sub-scores for physical, social, and self-presence, our analysis captures the effects of physicality across multiple dimensions of presence. This structure enables a more nuanced evaluation of how different levels of physical rendering influence not only overall presence, but also its constituent components.

\paragraph{Total Presence}
A significant main effect of physicality was observed for total presence ($F(1.51, 77.03)=47.34$, $p<0.001$, $\eta_p^2=0.48$). Post hoc Bonferroni-corrected comparisons indicated that the no-physicality (NP) condition yielded significantly lower total presence scores than both static physicality (SP; $M_{diff}=-0.72$, $SE=0.10$, $p<0.001$) and dynamic physicality (DP; $M_{diff}=-0.89$, $SE=0.12$, $p<0.001$). In addition, DP produced significantly higher total presence scores than SP ($M_{diff}=0.17$, $SE=0.07$, $p=0.049$). Neither the interaction between physicality and interaction type ($p=0.069$) nor the between-subject effect of interaction type ($p=0.585$) reached significance.

\paragraph{Physical Presence}
A significant main effect of physicality was found for physical presence ($F(1.48, 75.69)=29.83$, $p<0.001$, $\eta_p^2=0.37$). NP resulted in significantly lower physical presence scores than both SP ($M_{diff}=-0.61$, $SE=0.11$, $p<0.001$) and DP ($M_{diff}=-0.75$, $SE=0.13$, $p<0.001$). No significant difference was observed between SP and DP ($p=0.164$). Neither the interaction effect ($p=0.104$) nor the main effect of interaction type ($p=0.430$) was significant.

\paragraph{Social Presence}
A significant main effect of physicality was also observed for social presence ($F(1.95, 99.66)=28.48$, $p<0.001$, $\eta_p^2=0.36$). Social presence scores were significantly lower in the NP condition than in both SP ($M_{diff}=-0.61$, $SE=0.10$, $p<0.001$) and DP ($M_{diff}=-0.81$, $SE=0.12$, $p<0.001$). No significant difference was detected between SP and DP ($p=0.277$). No interaction effects or between-subject effects of interaction type were observed ($p=0.368$ and $p=0.863$, respectively).

\paragraph{Self-Presence}
Physicality also exerted a significant main effect on self-presence ($F(1.36, 69.56)=36.55$, $p<0.001$, $\eta_p^2=0.42$). Self-presence scores in the NP condition were significantly lower than those in both SP ($M_{diff}=-0.97$, $SE=0.16$, $p<0.001$) and DP ($M_{diff}=-1.18$, $SE=0.18$, $p<0.001$). The difference between SP and DP did not reach significance ($p=0.063$). No significant interaction effects or between-subject effects of interaction type were observed.

\subsubsection{Realism}
Levene’s test of equality of error variances indicated a violation for realism scores in the dynamic physicality (DP) condition ($p = 0.002$). Accordingly, Welch’s ANOVA was used to evaluate between-subject differences for the DP condition, in addition to the mixed ANOVA conducted across all physicality levels. A significant main effect of physicality was observed for realism ($F(1.59, 80.97)=29.40$, $p<0.001$, $\eta_p^2=0.37$). Post hoc Bonferroni-corrected comparisons showed that the no-physicality (NP) condition yielded significantly lower realism scores than both static physicality (SP; $M_{diff}=-1.26$, $SE=0.25$, $p<0.001$) and dynamic physicality (DP; $M_{diff}=-1.54$, $SE=0.23$, $p<0.001$). No significant difference was detected between SP and DP ($p=0.213$), nor were there significant between-subject effects of interaction type ($p=0.311$).

A significant interaction between physicality and interaction type was observed for realism ($F(3.18, 80.97)=2.81$, $p=0.042$, $\eta_p^2=0.10$), indicating interaction-specific patterns. Follow-up Bonferroni-corrected comparisons revealed that for object handovers, realism scores in the NP condition were significantly lower than those in both SP ($M_{diff}=-1.28$, $SE=0.43$, $p=0.012$) and DP ($M_{diff}=-1.67$, $SE=0.40$, $p<0.001$), with no significant difference between SP and DP ($p=0.527$). A similar pattern was observed for fist bumps, where NP again resulted in significantly lower realism scores than both SP ($M_{diff}=-1.78$, $SE=0.43$, $p<0.001$) and DP ($M_{diff}=-2.33$, $SE=0.40$, $p<0.001$), while SP and DP did not significantly differ ($p=0.114$). For high fives, evaluated using Welch’s ANOVA as noted above, no significant differences were observed among physicality conditions ($p=0.136$).

\subsubsection{Enjoyability}
A significant main effect of physicality was found for enjoyability ($F(1.49, 75.74)=16.21$, $p<0.001$, $\eta_p^2=0.24$). Post hoc Bonferroni-corrected comparisons revealed that NP produced significantly lower enjoyability scores than both SP ($M_{diff}=-0.91$, $SE=0.18$, $p<0.001$) and DP ($M_{diff}=-0.67$, $SE=0.20$, $p=0.004$). No significant difference was observed between SP and DP ($p=0.088$). Neither the interaction between physicality and interaction type ($p=0.224$) nor the between-subject effect of interaction type ($p=0.972$) reached significance.

\subsubsection{Comfort}
No significant main effect of physicality was observed for comfort ($p=0.114$), nor were there significant effects of interaction type ($p=0.899$) or an interaction between physicality and interaction type ($p=0.247$).

To further assess the practical equivalence of comfort across conditions, equivalence tests (TOST; margin $\pm0.5$ on the 7-point scale) were conducted. The NP–SP comparison ($M_{diff}=-0.333$, 90\% CI [–0.576, –0.091]) crossed the lower equivalence bound, indicating non-equivalence. In contrast, both the SP–DP ($M_{diff}=0.222$, 90\% CI [–0.015, 0.459]) and NP–DP ($M_{diff}=-0.111$, 90\% CI [–0.426, 0.204]) comparisons fell within the equivalence bounds, though neither reached statistical significance after Holm correction ($p=0.066$).

\subsubsection{Connection to Virtual Character}
A significant main effect of physicality was observed for perceived connection to the virtual character ($F(1.49, 75.96)=34.38$, $p<0.001$, $\eta_p^2=0.40$). Post hoc Bonferroni-corrected comparisons indicated that NP resulted in significantly lower connection scores than both SP ($M_{diff}=-1.50$, $SE=0.23$, $p<0.001$) and DP ($M_{diff}=-1.41$, $SE=0.24$, $p<0.001$). No significant difference was observed between SP and DP ($p=1.000$). Neither the interaction between physicality and interaction type ($p=0.231$) nor the between-subject effect of interaction type ($p=0.797$) was significant.

\section{Discussion}\label{sec:Discussion}
This work set out to examine whether ETHDs can support socially meaningful physical interactions in immersive virtual environments, and how different levels of physical rendering influence user experience. The results demonstrate that introducing true physical contact—whether static or dynamic—substantially improves experiential outcomes compared to purely virtual interaction. At the same time, the findings reveal important limitations in how current dynamic rendering strategies translate to perceived realism and comfort across different interaction types. In this section, we interpret these results in the context of encountered-type haptics and social–physical HRI, discussing what the observed benefits and shortcomings imply for the design of future systems.


\subsection{ETHD Feasibility for Social-Physical Interactions}
Accurate spatial alignment is a foundational requirement for ETHDs. If the physical prop does not sufficiently coincide with its virtual representation, interactions may feel unnatural or become infeasible. In ETHOS, spatial misalignment is dominated by the fiducial-based registration process, as the robotic manipulator itself provides high positioning accuracy. The resulting colocation error of $5.09 \pm 0.94$ mm falls well below the 1~cm threshold identified by Fremerey et al.~\cite{fremerey2020development}, indicating that static colocation errors are unlikely to be perceptible during interaction.


Interactions with ETHDs demand more than accurate spatial colocation; they also require contact to occur at the right moment. Delays can disrupt the sense of realism, making precise temporal alignment essential for convincing interactions. Across the subsystems of ETHOS, sources of delay such as message passing and hand-tracking latency can shift the timing of contact, creating misalignment between the user’s hand and the physical prop. Characterizing these temporal factors is therefore critical to assessing the suitability of our system for recreating natural interactions. Latency perception thresholds (LPTs) define the maximum tolerable delay in a system before user experience or task performance is degraded, and they have long been a focus of VR research. Factors such as task difficulty and interaction speed can significantly influence LPTs \cite{yang2025dynamic}, with the addition of haptic feedback further heightening sensitivity to delay \cite{lee2009discrimination}. As a result, defining a single latency perception threshold (LPT) for ETHD interactions is challenging. These interactions span multiple phases—from free-hand motion to physical contact—each of which aligns more closely with different perceptual thresholds reported in the literature. For example, an LPT derived from a virtual pointing task (130–170 ms) \cite{yang2025dynamic} differs substantially from one based on a tapping task with a haptic feedback device (50 ms) \cite{lee2009discrimination}. Despite this variability, the average interaction latencies measured in this work, for both static and dynamic tracking conditions, remained below all reported thresholds. However, when one standard deviation is considered, the most stringent LPT discussed here (50 ms) is exceeded in many trials. This suggests that while mean system performance is likely sufficient even under highly sensitive conditions, perceptible latency effects may still arise.

Taken together, the above metrics directly address RQ2 by quantifying the spatial and temporal performance of ETHOS. These results also inform RQ1, highlighting spatial alignment and temporal coordination as essential design considerations for enabling encountered-type social–physical interactions in immersive virtual environments. While not examined quantitatively in this work, the physical characteristics of the interaction props represent another critical design dimension. As the physical form and material properties of the props ultimately mediate how interactions are rendered and perceived, understanding how deviations between intended and realized physical interactions influence user experience remains an important direction for future investigation.

\subsection{Experiential Outcomes of Social–Physical Interaction}
A system’s feasibility cannot be established solely through technical soundness; it must also support meaningful user experiences. To this end, the user study complements the technical evaluation by examining how social–physical interactions are experienced in immersive virtual reality, using both quantitative measures and qualitative participant feedback.

\subsubsection{Effects of Physicality}
The three physicality conditions allowed us to evaluate social–physical interactions across increasing levels of embodiment: NP as a purely virtual baseline, SP as conventional ETHD implementation, and DP as our dynamic extension. Results indicate a clear separation between the virtual baseline and the physicality conditions, with both SP and DP producing improvements over NP. However, contrary to expectations, DP did not consistently extend the benefits observed with SP. Together, these findings address the influence of the degree of physicality rendered (RQ4), while qualitative participant quotes are used to further uncover how these interactions were experienced in immersive virtual reality (RQ3).

\paragraph{No Physicality}
The NP condition performed the worst across nearly all evaluated metrics. In terms of presence, both the total score and all sub-dimensions were significantly lower compared to conditions with added physicality (SP and DP). Given this work’s focus on social–physical interactions—experiences that are simultaneously physically engaging and socially meaningful—the poor performance of the purely virtual baseline underscores the importance of incorporating not only the character’s social behaviours, but also the realism of physicality within these interactions. This outcome is likely attributable to breaks in presence (BIPs) during the virtual interaction, during which attention shifts from the virtual experience to the physical world \cite{gonccalves2025unified}. Participants reported that “the lack of physical stimuli quickly removed the feeling of reality from the virtual space” and that the NP condition “reminded me that the simulation was not real,” reinforcing that unmet expectations of physicality can lead to BIPs that negatively impact experience. 

Beyond presence, realism in handover and fist-bump interactions also scored lowest in the NP condition. These poor realism scores further highlight the critical role of physicality in sustaining believable social–physical interactions. Consistent with the enjoyment and character connection results, participants described the purely virtual interaction as “unsatisfying” and “disappointing,” underscoring that the NP condition is insufficient to support meaningful social–physical interaction.

\paragraph{Static Physicality}
The SP condition represents a first step toward recreating the true physicality of the selected interactions. Here, the role of physical props is isolated, as each remains statically aligned while participants engage with the virtual character. Presence scores, both total and across all sub-dimensions, improved significantly compared to the NP condition. This finding aligns with prior work, such as Gibbs et al. \cite{gibbs2022comparison}, which demonstrated that introducing haptic feedback to a purely virtual baseline enhances presence in VR. Participant accounts help explain this effect, noting that the physical aspect of the interactions “helped ground me within the virtual environment” and that it “really felt like I was existing and interacting with the virtual environment” when engaging with the props.

Improvements in physical and social presence were expected, as tangible contact aligns with the intended form of these interactions. More notably, however, the increase in self-presence represents a particularly meaningful outcome. As outlined in Table \ref{tab:MPS}, self-presence encompasses bodily and emotional connectivity as well as the sense of bodily extension. Enhancements in this dimension suggest that physical rendering not only strengthens the authenticity of the interaction but also deepens the user’s perceived role within it. Enhancements in self-presence suggest that physical rendering not only strengthens interaction authenticity but also deepens the user’s perceived role within the experience. This increase in self-presence aligns with prior findings that multisensory integration (e.g. visuotactile feedback) strengthens embodiment subcomponents such as body ownership and agency \cite{gall2021embodiment}. In VR studies comparing a visual-only experience to one with added tactile feedback, participants show stronger illusions of being touched and higher embodiment ratings \cite{seinfeld2022evoking}, supporting the idea that physical rendering deepens users’ perceived role in interactions.

Significantly higher scores in realism, enjoyability, and character connection further support the improvement of experience within the SP condition as compared to NP. Participants emphasized this effect, noting that “the calibration [between virtual and physical worlds] was quite accurate, making the interaction feel real”, that “adding a physical component to my interaction with the character made our interaction much more enjoyable”, and that the physical contact with the prop "made me feel more connected to the virtual character." Taken together, these results demonstrate that the SP condition substantially improves the user experience and underscores the experiential feasibility of encountered-type haptics for recreating social–physical interactions (RQ1).

\paragraph{Dynamic Physicality}
As the selected social–physical interactions inherently involve non-static contact—such as the coordinated motion and timing required for object handover \cite{strabala2013toward}—the DP condition extends SP by introducing both motion and impact. While DP was expected to yield broader experiential improvements over SP, significant differences were observed only for total presence. This pattern suggests that dynamic rendering alone is insufficient to guarantee improved experience; rather, social–physical interactions are highly sensitive to control fidelity, temporal coordination, and interaction-specific dynamics. Participant feedback further reinforces this interpretation. Although some participants described DP interactions as “seamless” or “firm,” others reported experiences that felt “jarring” or “misaligned,” indicating that even subtle deviations in timing or motion quality can meaningfully alter perceived interaction quality. Importantly, this outcome should not be interpreted as a failure of dynamic physicality, but as a design insight for ETHDs. Specifically, it highlights that achieving convincing social–physical interaction requires not only acceptable mean spatial alignment and latency, but also interaction-specific tuning of dynamics and temporal behaviour. These findings therefore directly inform RQ1, further outlining key design considerations beyond simple technical performance.

One possible explanation for this variability lies in the control and detection strategies employed in the current implementation. In particular, a common trajectory generation strategy was used across all interaction types \cite{FastHandovers}, rather than tailoring trajectories to the unique dynamics of each interaction. This design choice likely reduced robustness and perceived naturalism across the interaction set. Additionally, interaction completion was detected using simple force thresholds, which may have produced unnatural sensations during more complex contact phases of dynamic interaction. Prior work, such as Chan et al. \cite{chan2012grip}, highlights the importance of modelling transient grip forces in human–human handovers, suggesting that discrete thresholds for contact recognition are insufficient to capture the timing and compliance required for natural interaction rendering.

\subsubsection{Effects of Interaction Type}
The social–physical interactions evaluated in this work were object handovers, fist bumps, and high-fives. To the authors’ knowledge, this work represents the first application of social–physical HRI within VR and the first experiential evaluations of these interactions using an ETHD approach.
Across the interaction types, no main effects on presence, enjoyability, comfort, or avatar connection were found. This suggests that the level of physicality rendered is the primary determinant of experience quality, and that the experiential improvements produced by ETHOS are broadly applicable across interactions that share common elements but differ in their patterns of contact and motion.

Realism, however, was an outlier in this trend, revealing an interaction-specific effect: while handover and fist bump benefited from added physicality, the high five interaction did not vary across conditions. Participant feedback indicated that the silicone hand prop lacked sufficient fidelity, limiting perceived realism despite other metrics improving. This highlights the importance of prop design and suggests that meaningful experiential gains are possible even without perfect physical replication.

\subsubsection{Comfort Across Conditions}
Comfort scores did not differ significantly across physicality conditions, and practical equivalence could not be established through TOST analysis. These findings suggest that comfort remained broadly consistent across interaction types, though subtle differences cannot be ruled out. This outcome motivates further investigation into user comfort within immersive experiences involving encountered-type physical interactions. As comfort was assessed using a single-item measure, future work should consider a more detailed evaluation of comfort subcomponents (i.e., physical strain, psychological ease) to better identify potential effects.

\subsection{Insights and Limitations}
ETHOS represents the initial implementation of an ETHD for enabling social–physical interactions in VR. As an experimental, proof-of-concept platform, several system-level simplifications were undertaken in its current design. For example, the physical props were designed to approximate real-world counterparts but were not formally characterized, a dynamic interaction control strategy was adapted from object handovers and applied across all interaction types, and force thresholds were heuristically defined from preliminary pilot observations without systematic justification or consideration of torques. Future work should address these limitations by characterizing prop materials, developing interaction-specific control strategies, and determining representative force–torque profiles. In addition, manipulator compliance should be explored to better reflect the natural compliance of the human arm across the phases of the developed interactions. The observation of positive experiential outcomes despite these system-level simplifications highlights the potential for even greater experiential benefits as these limitations are addressed in future work.

A key takeaway from user feedback is the complexity of recreating interpersonal interactions with virtual humans. Readiness and greeting cues (e.g., gaze, waving, eye contact) were effective in guiding participants within this work, yet aspects of the virtual character’s appearance and animation were described as “lifeless” or “odd,” underscoring the sensitivity to realism in virtual interpersonal contexts. This aligns with Guadagno et al. \cite{guadagno2007virtual}, who found that behavioural realism is correlated with social presence. Future iterations of ETHOS should therefore prioritize enhancing the behavioural realism of the virtual character to foster stronger connections and more meaningful interactions. Verbal cues were also omitted in this implementation, necessitating further investigation into their addition in future work.

Another avenue to improve the ETHOS experience is to increase user agency and incorporate system adaptations tailored to individual characteristics. At present, ETHOS is constrained to a fixed interaction sequence with a predefined end goal, limiting user control. Agency (i.e., the perceived ability to influence elements of the virtual environment) has been shown to positively affect presence \cite{agency}. Extending ETHOS to support additional non-physical interactions or to allow users to guide the flow of interaction could therefore enhance presence and deepen engagement. Participant feedback also emphasized the value of personalization, with some noting that character proximity felt “too close” or that interaction height was not “ideal.” Unlike natural human–human encounters, which involve continuous mutual adjustment \cite{strabala2013toward}, ETHOS currently lacks sufficient adaptive and personalized behaviours, suggesting an important direction for future development.

\section{Conclusion}\label{sec:Conclusion}
This work introduced ETHOS, an encountered-type haptic display designed to support socially meaningful physical interaction in immersive virtual reality. By integrating a torque-controlled robotic manipulator with interchangeable physical props and fiducial-based physical–virtual coordination, ETHOS enables social–physical interactions with a virtual character, including object handover, fist bump, and high five.

Technical characterization demonstrated that ETHOS achieves sub-centimetre spatial colocation and interaction latencies that remain, on average, below reported perceptual thresholds, establishing the feasibility of spatially and temporally aligned contact even under dynamic tracking. A subsequent 55-participant user study revealed that purely virtual interaction is insufficient to sustain believable social–physical experience. Introducing physical contact through encountered-type haptics significantly enhanced presence, realism, enjoyability, and perceived connection to a virtual character. While dynamic physicality produced a modest increase in total presence over static physicality, it did not consistently improve other experiential measures as expected, highlighting the sensitivity of social–physical interaction to timing, compliance, and interaction-specific dynamics. Rather than indicating a limitation of dynamic physicality itself, these findings underscore the need for more tailored motion strategies and adaptive force control when rendering interpersonal touch.

Taken together, these results establish the technical and experiential feasibility of encountered-type haptics for supporting socially meaningful physical interaction in immersive virtual reality. Beyond feasibility, this work connects social–physical HRI, encountered-type haptics, and immersive VR by examining how the coordination of physical and virtual subsystem design shapes user experience. The findings indicate that physical contact plays an important role in supporting presence and perceived interpersonal connection, and that constraints in one subsystem can directly influence experiential outcomes in the other. Future work will build on these insights through interaction-specific trajectory generation, compliance-aware contact modeling, and richer virtual character behaviours.

\bibliographystyle{ACM-Reference-Format}
\bibliography{sample-base}


\end{document}